# Understanding Brownian yet non-Gaussian diffusion via long-range molecular interactions.

Francisco E. Alban-Chacón[a*], Erick A. Lamilla-Rubio[b,c] & Manuel S. Alvarez-Alvarado[d]

a Faculty of Natural Science and Mathematics, Escuela Superior Politécnica del Litoral, EC090112 Guayaquil, Ecuador, alban@espol.edu.ec

b Faculty of Natural Science and Mathematics, Escuela Superior Politécnica del Litoral, EC090112 Guayaquil, Ecuador, ealamill@espol.edu.ec

c Facultad de Ciencias Matemáticas y Físicas, Universidad de Guayaquil, 090514 Guayaquil, Ecuador, erick.lamillaru@ug.edu.ec

d Faculty of Electrical and Computer Engineering, Escuela Superior Politécnica del Litoral, EC090112 Guayaquil, Ecuador, mansalva@espol.edu.ec

* Corresponding author e-mail address: alban@espol.edu.ec

## Abstract

In the last years, a few experiments in the fields of biological and soft matter physics in colloidal suspensions have reported "normal diffusion" with a Laplacian probability distribution in the particle's displacements (i.e., Brownian yet non-Gaussian diffusion). To model this behavior different stochastic models had been proposed, with all of them introducing new random elements that incorporate our lack of information about the media. Although these models work in practice, due to their own nature a thorough understanding of how the media interacts with itself and with the Brownian particle in Brownian yet non-Gaussian diffusion is outside of their aim and scope. For this reason, a comprehensive mathematical model to explain Brownian yet non-Gaussian diffusion that includes molecular interactions is proposed in this paper. Based on the theory of interfaces by Gennes and Langevin dynamics, it is shown that long-range interactions in a weakly interacting fluid and in a microscopic regime of zero viscosity leads to a Laplacian probability distribution in the particle's displacements. Further, it is shown that a phase transition can explain a high diffusivity and causes this Laplacian distribution to evolve towards a Gaussian via a transition probability in the interval of time as it was observed in experiments. To validate these model predictions, the experimental data of the Brownian motion of colloidal beads on phospholipid bilayer by Wang *et al.* is used and compared with the results of the theory. This comparison suggests that the proposed model not only is able to explain qualitatively the Brownian yet non-Gaussian diffusion, but also quantitatively.

# I. Introduction

The dynamics of Brownian diffusion is frequently used for modeling stochastic motions to get information about the particle's interaction with binding partners and the local environment [1], [2]. The main characteristic of Brownian diffusion lies in the feature of random wiggling particle motion that generally produces a normal or Gaussian distribution in the particle density function, with mean $\mu = 0$ and variance $\sigma^2 = 2Dt$; where $D$ is interpreted as mass diffusivity or diffusion coefficient [3], [4]. The Brownian diffusion model is very useful to analyze and study a variety of physical processes related to mechanisms of particle transport [5], [6], thermal fluctuations [7], [8], [9], particle manipulation [10], [11], [12] and stellar dynamics [13], [14]. Despite of this gaussian behavior being common in the displacement distribution for systems that exhibit Brownian motion, in the last years, efforts had been made to study a new type of Brownian diffusion. Like normal diffusion, it has a linear time dependence of the mean-square displacement (MSD), but is accompanied by a non-gaussian displacement distribution, which had been identified as "anomalous yet Brownian" diffusion [15]. The Brownian yet non-Gaussian diffusion reported by Wang *et al*. is based in the classical random walk in which mean-square displacement is simply proportional to time but, instead, has an exponential behavior with the decay length of the exponential being proportional to the square root of time. This concept was vastly used to propose a model based in anomalous, but Brownian diffusion to describe the nature of diffusivity memory, but not the memory in the direction of the particle's trajectories. The model was coined as diffusing diffusivity, due to the random walk that the diffusivity experiences [16]. Chubinsky-Slater's idea in [16] had also been useful to model the behavior of biological, soft, and active matter systems establishing a minimal model framework of diffusion processes with fluctuating diffusivity [17]. Also, in the realm of fluids, it had been shown that in some confinement conditions, density fluctuations might be relevant to Brownian yet non-Gaussian diffusion [18], [19].

The literature presents different kind of stochastic models that explain mathematically the Brownian yet non-Gaussian diffusion process. This includes: studies of the role of media heterogeneity by randomizing parameters that appear in diffusivity dynamical equations [20]; demonstrations on time-dependent diffusivity, induced by external non-thermal noise [21] and interesting comparisons between non-gaussian random diffusivity models [22]. All the exposed models have in common that they reflect different levels of our lack of information/ignorance about the surroundings/media. The lack of information about the media is considered by introducing random elements to the models via random parameters, random diffusivity, or noise.

From a practical point of view, there is plenty of evidence that these models give correct predictions. Nevertheless, from a theoretical perspective to the best of our knowledge a physical mechanism that incorporate media information a priori by establishing a link between deterministic molecular interactions and Brownian motion had not been studied. This fact motivates the study and validation of a model that considers long-range Brownian particle-fluid molecular interactions, which via a phase transition (in a compressible fluid) attempts to explain the Brownian yet non-Gaussian diffusion by providing accurate predictions. (Including the experimental observation of the transition to a Gaussian process as observed in [15]).

The rest of the paper is structured as follows: Section II presents the theoretical/mathematical framework that explains how a Brownian particle interacts with a fluid. Section III discusses relevant experimental information and how it compares to the proposed model. Section IV provides experimental data and quantitively evaluate the predictions of the model. Finally, Section V incorporates the conclusion.

## II. Mathematical model of an interacting Brownian particle

To explain the results of recent experiments [15], [23], [24] and [25] a model that considers molecular interactions during Brownian motion is proposed as follows. Let's start with three fundamental assumptions of the model:

- There are two possible regimes: a microscopic scale and a macroscopic scale. As stated in [26], the microscopic scale is the one where the liquid can be treated as individual particles interacting between them and undergoing thermal motion. The macroscopic scale is the one where the fluid satisfies Navier-Stokes equation. In [26] is determined that in the microscopic scale shear-stress viscous friction is zero and the liquid only has elastic properties. Friction emerges abruptly in the macroscopic scale, where the liquid can be treated classically.

- In the microscopic regime the Brownian particle interacts effectively with the molecules of the liquid via long-range interactions. (i.e., Van der Waals interactions). Via a Lennard-Jones attractive potential energy of the form [27]:

$$(1)\ U(r) \propto -\left(\frac{\sigma}{r_{ij}}\right)^6,$$

    Where $\sigma$ is the Van der Waals radius (i.e., known as the "size of the particle") and $r_{ij}$ is the distance between the Brownian particle and a fluid molecule.

- At all times the noise acting on the Brownian particle (sphere) is white, and Gaussian and its mean squared displacement is:

$$(2)\ \langle r^2 \rangle \propto t,$$

From the two initial assumptions above there are two possible outcomes:

1. If the Brownian particle interaction energy with the molecules of the liquid is greater than the energy required to break the intermolecular forces at some external pressure P and temperature T, the liquid behaves as a Van der Waals gas (Analogous to the behavior of water in solution with strong dissociated ions like $Na^+$ and $Cl^-$ [28]). As can be deduced from (1), this occurs in the microscopic regime up until some distance, where the interaction energy of the Brownian particle-fluid has decreased enough so that the Van der Waals gas start behaving as a liquid again (Note that this could take a very long time or could be relatively fast. In any case this should be determined experimentally). This gives rise to the macroscopic regime, and it can be seen as a phase transition.

2. If the Brownian particle interaction energy with the molecules of the liquid is less than the energy required to break the intermolecular forces, then no phase transition occurs, the microscopic scale becomes inaccessible, and the liquid behaves classically for all times (i.e., fully incompressible liquid). Meaning Brownian motion as described by Einstein takes place. Note that this will typically be the case since the interaction energies of molecules in liquids are typically high and most liquids are treated as incompressible. For example, water molecules exhibit hydrogen bonding, which are generally stronger than Van der Waals interactions [28].

It is outcome 1 that originates new physics and is the case of interest in this paper. Is important to mention that as usual only self-interactions and pair-wise interactions [29] via fluid between Brownian particles will be considered. Also, although the derivation will be performed for 3-D it holds for 1-D and 2-D. The derivation goes as follows:

The long-range fluid particles-Brownian particle interactions stated in 1. are modeled as in vapor-liquid interfaces as had been previously realized by Gennes [30]. In this situation the fluid is an ideal gas hitting a spherical interface interacting via an effective long-range Lennard-Jones type potential energy given in (1). The Van der Waals radius $\sigma$ is given by:

$$(3)\ \sigma = \frac{D+d}{2},$$

where, $D$ is the diameter of the Brownian particle and $d$ is the diameter of the fluid particle. For this case $d \ll D$, then:

$$(4)\ \sigma = R,$$

being $R$ the radius of the Brownian particle. The potential derived from (1) is:

$$(5)\ V(r) \propto -\frac{1}{r^3},$$

using Boltzmann-Gibbs distribution and assuming that $\rho_L \ll \rho_{bp}$ (i.e., density of the liquid and density of the Brownian particle), then Gennes [30] arrived at:

$$(6)\ \frac{\rho(r)-\rho_L}{\rho_s-\rho_L}=\left(\frac{R}{r}\right)^3.$$

Originally $\rho(r)$ was a function of $z$, which measured the height from the interface. $z$ has been replaced by $r$ which measures the radial distance from the center of the Brownian particle and of course indicates that there is spherical symmetry. Also, the $\rho_L$ and $\rho_s$ terms are there to satisfy the boundary conditions. When $r \to \infty$, then $\rho_L$ is recovered as one should expect. When $r = R$ the density of the fluid at the interface is recovered and is called $\rho_s$.

The relevant density differences in (6) are $\tilde{\rho} = \rho(r) - \rho_L$ and $\rho_b = \rho_s - \rho_L$. Equation (6) is:

$$(7) \quad \tilde{\rho} = \rho_b \left(\frac{R}{r}\right)^3.$$

Since there is radial symmetry, equation (7) can be further simplified, given that:

$$(8) \quad \tilde{\rho} = \frac{dN}{dV} = \frac{dN}{dr} \frac{1}{(4\pi r^2)},$$

$$(9) \quad \rho_b = \frac{1}{(4\pi R^2)} \frac{dN}{dr}\Big|_{r=R},$$

notice that $\frac{dN}{dr}\Big|_{r=R} = -\frac{1}{R}$. Therefore,

$$(10) \quad \frac{dN}{dr} = -\left(\frac{1}{r}\right),$$

Equation (10) was obtained by treating the fluid as an ideal gas, as was done by Gennes in a different physical situation [30]. However, it is required in this model to consider molecular interactions in the gas (Van der Waals gas). This type of gas can be treated as an ideal gas with a modified number of particles $ZN$. Where Z is the compressibility factor [31]. $Z < 1$ means attractive interactions, Z=1 means no interaction (ideal gas) and Z>1 means repulsive interactions between fluid molecules. Then, equation (10) simply becomes:

$$(11) \quad \frac{d(ZN)}{dr} = -\left(\frac{Z}{r}\right),$$

It is this gradient of the number of particles that generates a radial attractive interaction between two Brownian particles. Also, since the derivation of (11) was performed in the realm of classical statistical mechanics, to indicate that $r$ is not a random variable will be replaced by $\langle r \rangle$, which represents a mean radial distance. Note that the number of particles gradient in (11) exists only in the radial direction. Further, the difference $d(ZN)$ is taken with respect to the number of particles the fluid would have in absence of the Lennard-Jones potential.

Also, the number of particles gradient in (11) do not generate any force acting on the same Brownian particle that is producing it (self-interaction). Of course, this is due to its spherical symmetry. However, it does generate a force acting on a different Brownian particle (pair-wise interaction). This force is derived as follows:

At any given time, consider Brownian particle 1 located a mean radial relative distance $\langle r \rangle$ from Brownian particle 2. (In the following, it will be assumed that one is in the reference frame of Brownian particle 2). As it is done in the derivation of mean free path with ideal gases [32] and [33] the Brownian particle will be treated as a point particle with cross section of $\pi R^2$, where R is the radius of the sphere. Now, imagine a cylinder enclosing Brownian particle 1 cross section oriented in the radial direction with the same cross section as the Brownian particle and height $d\langle r \rangle$ (see figure 2). This cylinder will have a greater number of particles $ZN_1$ in its farthest half (to Brownian particle 2) than in its closest half $ZN_2$ (see figure 1(b)). This means that the force acting on Brownian particle 1 will be simply due to the difference in number of particles and the ideal gas law (Considering Z) and will be acting radially inwards towards Brownian particle 2, as it can be seen from figure 1(a) below:

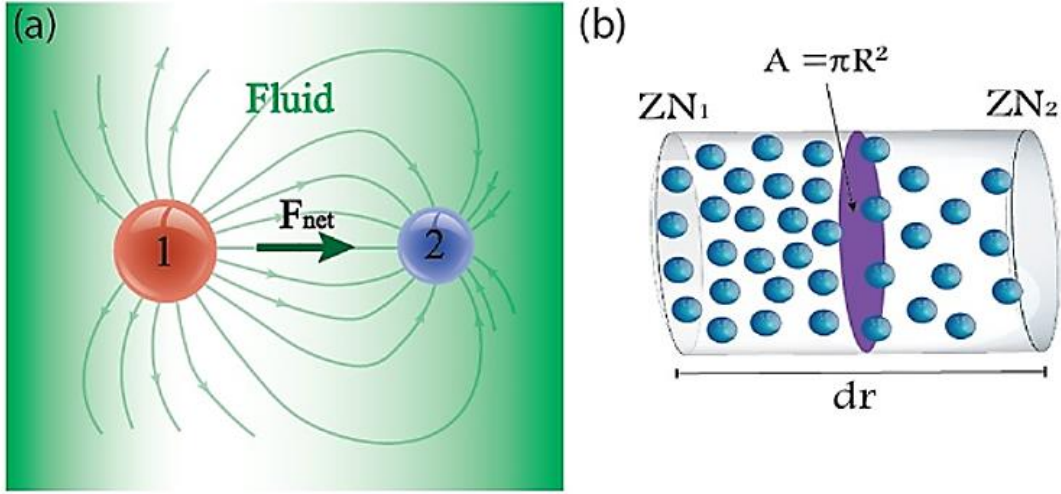

FIGURE 1

(a). Pair-wise interaction of 2 Brownian particles via the radial force field in (17). (b).Volume element that shows the gradient in the number of fluid particles (fluid particles in blue and cross section of Brownian particle 1 in purple) that generates a force acting on Brownian particle 1.

The magnitude of the force in figure 1(a) is given by:

$$(12) \quad F_n = \frac{Z(|N_2 - N_1|)}{d\langle r \rangle} k_B T = \frac{k_B T}{\left|\frac{d\langle r \rangle}{d(ZN)}\right|},$$

Now, using equation (11)

$$(13) \quad \left|\frac{d\langle r \rangle}{d(ZN)}\right| = \frac{\langle r \rangle}{Z},$$

Note, that $\langle r \rangle$ is not the root mean squared displacement, but rather the root mean squared radial relative distance. Then:

$$(14) \quad \frac{\langle r \rangle}{Z} = \frac{\sqrt{2}\sqrt{2Dt}}{2*Z} = \sqrt{\left(\frac{D}{Z^2}\right)t},$$

The root mean squared radial displacement of Brownian particle 1 is $\sqrt{2Dt}$ (There is only one degree of freedom, therefore it is not dependent on dimension). Since the relative displacement to Brownian particle 2 is of interest, then a factor of $\sqrt{2}$ is included in (14) [32]. Furthermore, the distance traveled as measured from Brownian particle 2 is required, therefore a factor of $\frac{1}{2}$ is also included in (15).

Finally, using (12), (13) and (14) the magnitude of the force is:

$$(15) \quad F_n = \frac{k_B T}{\langle r \rangle} = \frac{k_B T}{\sqrt{\left(\frac{D}{z^2}\right)t}},$$

From equation (15), $\left(\frac{D}{z^2}\right)$ can be seen as a new effective diffusion constant $D_{eff}$ that had changed due to molecular interactions of the fluid. Then:

$$(16) \quad D_{eff} = \left(\frac{D}{z^2}\right),$$

The force that corresponds to (15) is given by (considering the negative sign in (11)):

$$(17) \quad \vec{F} = -\frac{k_B T}{\sqrt{D_{eff} t}} \hat{r},$$

This force can be included in Newton's second law starting from the Langevin approach as follows:

$$(18) \quad m\ddot{\vec{r}} = -\alpha \dot{\vec{r}} - \frac{k_B T}{\sqrt{D_{eff} t}} \hat{r} + \varepsilon(t),$$

The first term to the right of the equal sign represents the viscous drag force generated by the fluid (with $\alpha$ being the drag coefficient), the second term is the density gradient force introduced in this model and the third term $\varepsilon(t)$ is the white gaussian noise generally assumed in Brownian Motion. In the Brownian regime the inertial term in the left-hand side of the equation is zero and in the microscopic regime assumed in this model, the viscous term is zero as well. Then, the simplified equation in the radial direction is:

$$(19) \quad 0 = -\frac{k_B T}{\sqrt{D_{eff} t}} + \varepsilon(t),$$

The Fokker-Planck equation derived from (19), treating the force in (17) as a constant term in the position $\vec{r}$ (i.e., this term acts as a dry friction term in the position [34]) is derived in [35]. Since the stationary solution is needed, the Boltzmann-Gibbs distribution is the solution [35]. The argument of the Boltzmann-Gibbs distribution being the potential derived from the force in (17), which is:

$$(20) \quad V(r) = \frac{k_B T}{\sqrt{D_{eff} t}} |\vec{r}|,$$

And the Boltzmann-Gibbs distribution:

$$(21) \quad p(\vec{r}) = A e^{-\frac{V(r)}{k_B T}},$$

Where $A$ is a constant. Note that the potential is linear in the absolute value of the position. The full stationary probability distribution is therefore (Note that the $k_B T$ term cancels):

$$(22) \quad p(\vec{r}, t) = \frac{1}{\sqrt{4 D_{eff} t}} e^{-\frac{|\vec{r}|}{\sqrt{D_{eff} t}}},$$

The expression in (22) is valid from 0 up to $t_c$. Where, as mentioned at the beginning of this section, there is some $\langle r \rangle_c$ or correspondingly some $t_c$ (should be determined experimentally and could be a very large value) at which the microscopic regime ends abruptly, and the macroscopic regime starts. The fluid stops acting as a gas and starts behaving as a liquid. In general, in these types of phase transitions at low pressures and far from the critical point the compressibility factor Z suffers a vast discontinuity [31]. It generally jumps from a value close to 1 to a very small value [31] that is reminiscent of a liquid. Of course, this indicates that interactions between molecules in the liquid phase are a lot stronger than in the gaseous phase. Now, since Z has a very large decrease, $D_{eff}$ has a very large increase. By noting that in this regime, the viscous drag force must be considered (as specified in the assumptions of this model), it can be realized that the magnitude of the force in (17) will be several orders of magnitudes less (high $D_{eff}$ )than the viscous drag force term in (18). Meaning, for this regime the force in (17) can be safely ignored. After $t_c$, the Langevin equation becomes:

$$(23) \quad \alpha \dot{\vec{r}} = \varepsilon(t'),$$

Or the corresponding Fokker-Planck equation being the diffusion equation [36]:

$$(24) \quad \frac{\partial p_{(\vec{r}, t')}}{\partial t'} = D_{effL} \nabla^2 p(\vec{r}, t'),$$

Which should be solved with (22) evaluated at $t_c$ as initial condition and $t' = t - t_c$. To account for the discontinuity of Z, the effective diffusivity for this regime will be denoted by:

$$(25) \quad D_{effL} = \frac{D}{Z_L^2},$$

And the effective diffusivity in the gaseous phase will be denoted by:

$$(26) \quad D_{effG} = \frac{D}{Z_G^2},$$

Since $D_{effG} \ll D_{effL}$, equation (22) can be seen as a delta function when evaluated at $t_c$ and used as an initial condition to solve (24). Meaning that after $t_c$ the probability density will have some transition probability that will evolve rapidly towards a Gaussian, which is:

$$(27) \quad p(\vec{r}, t') \sim \frac{1}{(4\pi D_{effL} t')^{\frac{3}{2}}} e^{-\frac{|\vec{r}|^2}{4 D_{effL} t'}},$$

From now on the probability distribution in (22) will be denoted as $p_G(\vec{r},t)$ (gaseous phase) and the full solution to (24) with the initial condition being $p_G(\vec{r},t_c)$ will be denoted as $p_L(\vec{r},t')$ (liquid phase).

The description above gives an accurate time evolution of the probability distribution of the position of the Brownian particles at two different scales (The microscopic scale with $D_{effG}$ and the macroscopic scale with $D_{effL}$). However, it does not tell you which probability distribution you should use at any given time (over a single scale over a time interval t) $p_G(\vec{r},t)$ or $p_L(\vec{r},t')$. Meaning at time $t < t_c$ a Brownian particle may be located at a position $\vec{r}$ such that it is either in the gaseous phase or in the liquid phase. The same happens at $t > t_c$. This means that the gaseous and liquid phases can be regarded as two possible states of the system, with no preference over one or the other. Of course, this should be accounted for when taking averages of observables, one should take the average over all possible states. Therefore, the mean squared displacement for this process will be given by:

$$(28) \quad \langle r(t)^2 \rangle = 0.5 \left( \int_{-\infty}^{\infty} |\vec{r}|^2 p_G(\vec{r},t)\, d\vec{r} + \int_{-\infty}^{\infty} |\vec{r}|^2 p_L(\vec{r},t)\, d\vec{r} \right),$$

Which simplifying gives:

$$(29) \quad \langle r(t)^2 \rangle = 0.5(6D_{effG}t + 6D_{effL}t),$$

Rearranging terms:

$$(30) \quad \langle r(t)^2 \rangle = 6[0.5(D_{effG}+D_{effL})t],$$

Where $D_{avg}$ can be seen as a diffusivity and is given by:

$$(31) \quad D_{avg} = 0.5(D_{effG}+D_{effL}),$$

Equation (30) can be rewritten as:

$$(32) \quad \langle r(t)^2 \rangle = 6D_{avg}t,$$

Equation (32) leads to interpret the whole Brownian process discussed until now with two different scales with diffusivities $D_{effG}$ and $D_{effL}$ as a single re-scaled process, such that the standard deviation of the PDF is given by (32). Thereby, from (32) and the analysis up until before equation (28), it can be inferred that the observed probability distribution of the position $\vec{r}$ of the Brownian particle will be given by:

$$(33) \quad p(\vec{r},t) = \begin{cases} p_{G'}(\vec{r},t'), & t < t_c \\ p_{L'}(\vec{r},t'), & t > t_c \end{cases},$$

with the probability distributions $p_{G'}(\vec{r},t)$ and $p_{L'}(\vec{r},t')$ being $p_G(\vec{r},t)$ and $p_L(\vec{r},t')$ re-scaled such that their mean squared displacement (i.e., standard deviation squared) at any given time is given by (32).

## III. Initial considerations

A few initial remarks concerning the proposed model and some relevant experimental observations are summarized here. In experiment [15] the Brownian motion of colloidal beads on phospholipid bilayer (DLPC) tubes (1-D) was studied. Lipid's bilayers tend to have weak intermolecular interactions between "lipids tails" [37]. (London dispersion forces are the type of molecular interactions occurring between lipids). The weaker these intermolecular interactions, the more flexible are the bilayers (i.e., membranes) and vice versa [37]. This means that the Brownian particle-fluid long-range interactions should be enough to cause a phase transition and the whole machinery developed above can be applied.

First, note that what is being measured in the experiment is $D_{avg}$, meaning:

(34) $$D_{avg} = 0.5\left(\frac{D}{Z_G^2} + \frac{D}{Z_L^2}\right),$$

Approximating the lipids tails in the gaseous phase as an ideal gas (i.e., almost no interaction), meaning $z_G \sim 1$, then:

(35) $$D_{avg} = 0.5(D + D_{effL}),$$

As mentioned before, $Z_L$ is a small number. This means $D_{avg} \gg D$, which is what was found in the experiment [15]. (i.e., unusually high diffusivity).

The Lennard-Jones potential introduced in this model causes a reduction of local pressure in the liquid. (Meaning the local pressure is less than the external pressure at which the experiment is being conducted). This means that determination of $Z_L$ requires full knowledge of the effective local pressure experienced in the liquid at the phase transition point, which is not an easy task to estimate theoretically and therefore a direct calculation of $D_{avg}$ is not performed here. (However, it could in principle be done). In any case, the value of $D_{avg} = 0.40 \ um^2 \cdot s^{-1}$ and $D$ reported in this experiment [15] can be used to calculate the value of $Z_L$ and check whether it is a reasonable value for a liquid at room temperature, low pressures and far from the critical point. D is calculated by using Einstein's relation:

(36) $$D = \frac{k_B T}{6\pi \eta_e R},$$

where T is temperature, $k_B$ is Boltzmann constant, R is the radius of the Brownian particle and $\eta_e$ is the extensional viscosity of the fluid. Extensional viscosity considers shear viscosity and bulk viscosity, which should be the case for a compressible fluid [38] and [39]. For a Newtonian fluid $\eta_e$ is given by [39]:

(37) $$\eta_e = 3\eta,$$

Where, $\eta$ is the shear viscosity of the fluid. Using (36) and (37):

(38) $$D = \frac{k_B T}{18\pi \eta R},$$

## IV. Model validation

In order to validate the proposed approach, experimental data taken from Wang *et al* [15] is employed. In this experiment, the temperature is reported to be $T = 22°C$, the viscosity of the media $\eta$ is reported to be $\approx 100$ times higher that than of bulk water and the radius of the Brownian particle is $R = 50\ nm$. Therefore, with the conditions of this experiment equation (38) is used to find that $D \sim 0.014\ um^2 \cdot s^{-1}$. Further, using this result and equation (25) it is found that $D_{effL} \sim 0.79 um^2 \cdot s^{-1}$ and $Z_L \sim 0.13$. (Note that $Z_L$ is a reasonable value for a liquid at room temperature and low pressures [31]). Also note that $D_{avg} \gg D$, since $0.40 \gg 0.014$.

Regarding the temporal evolution of the probability distribution, in this experiment $t_c \sim 4s$. Meaning up until $t = 4s$ the probability density function (PDF) in the displacement $x$ will be given by the analogous 1D expression to (22) with a diffusivity of $D_{avg}$. Numerical implementation of the proposed model in Mathematica shows that after $\sim 2s$ the PDF (22) transitions towards a 1D Gaussian in the displacement $x$ (see figures below). This is precisely what was observed in this experiment [15].

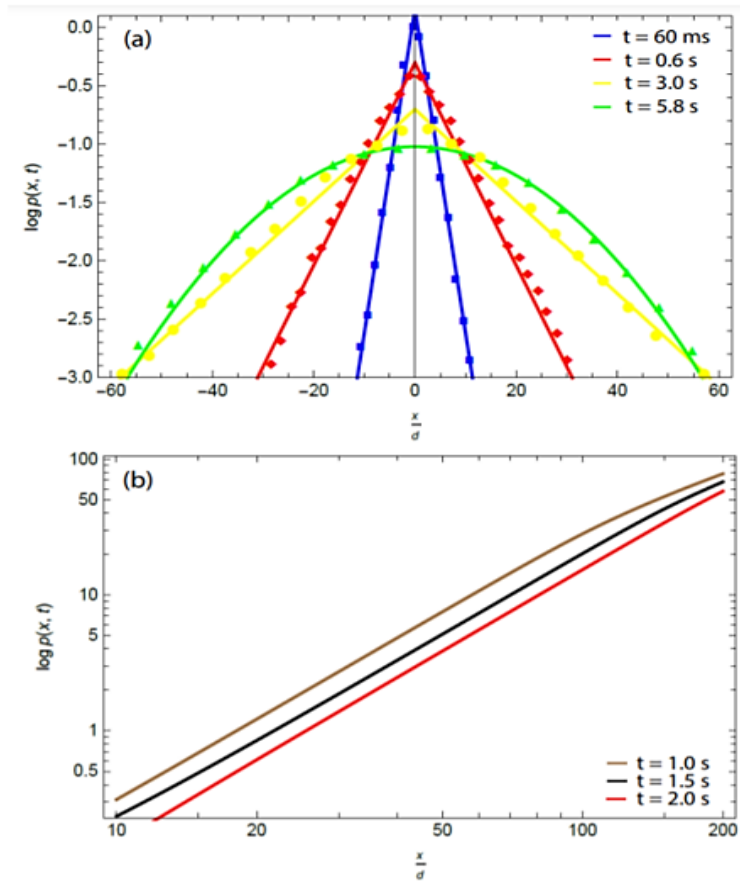

FIGURE 2

(a). Solid lines curves obtained from the model and experimental data points of the logarithm of the displacement probability distribution plotted against particle displacement normalized by the particle diameter d at different times (b). Theoretical prediction of the logarithm of the displacement probability distribution plotted against particle displacement normalized by the particle diameter d in a log-log scale at different times.

A brief description of the figures and data is given here. The solid lines curves of figure 2 (a) above were obtained from the proposed model. At $t = 60ms$, $t = 0.6\ s$, $t = 3s$, the corresponding 1D unnormalized equation (22) was used. It was re-scaled such that its mean square displacement is given by (32) 1D analogous (i.e., $\langle r(t)^2 \rangle = 2D_{avg}t$) with $D_{avg} = 0.40\ um^2 \cdot s^{-1}$.

At $t = 5.8s$, the differential equation (24) in 1D with $D_{effL} \sim 0.79 um^2 \cdot s^{-1}$ was solved numerically in Mathematica. The initial condition was taken to be the 1D unnormalized equation (22) evaluated at time $t = 4s$ with diffusivity $D_{effG} \sim 0.014\ um^2 \cdot s^{-1}$. The solution was re-scaled such that $\langle r(t)^2 \rangle = 2D_{avg}t$. As a result, the transition probability $p(x,t)$ at $t = 5.8s$ is found to be:

$$(39)\quad p(x) \propto 0.5e^{35.55-0.45x}\{erfc(5.96 - 0.038x) + e^{0.91x}erfc(5.96 + 0.038x)\},$$

Where:

$$(40)\ erfc(z) = 1 - erf(z),$$

With respect to the experimental data points, they were obtained from [15]. Linear regression analysis was performed to fit the experimental data points in figure 2 (a). By fixing the lowest order parameter, a calculation of the percentage error of the highest order coefficient/parameter between the theoretical prediction and the experimental best fit was performed at the corresponding times. At $t = 60ms$ the percentage error was found to be 0.28%, at $t = 0.6\ s$ the percentage error was found to be 2.0%, at $t = 3\ s$ the percentage error was found to be 2.0% and at $t = 5.8\ s$ the percentage error was found to be 3.4%.

By inspecting figure 2(b) it can be inferred that at some point before $200d$ the slope of the brown and gray lines changes. This indicates exponential decay for $t = 1s$ and $t = 1.5s$. At $t = 2s$. The slope of the red line is constant, which indicates Gaussian behavior for all $x$ up to $x = 200d$.

From the figures and table above, it can be concluded that the theoretical model agrees with experiment. Percentage error in the parameters do not exceed 3.5 %. Further, besides the theoretical prediction of the emergence of a Laplacian distribution and an eventual Gaussian distribution in the particle displacements, it is remarkable that the transition probability at $t = 5.8\ s$, and the transition time (which emerge from ideal scenarios) are an accurate representation of the experimental data.

Finally, as a second part of the experiment, the membranes of the fluid were filled with cholesterol. Everything else was held constant and no exponential distribution was observed. As it had been observed before, cholesterol provides rigidity to the membranes by strengthening molecular interactions between lipids [40]. Since not enough energy is provided to break those interactions by the interaction Brownian particle-fluid (via the Lennard-Jones potential proposed in this model), then no phase transition occurs, and one is in the regime of "regular Brownian motion". Further, in this case the diffusivity was observed to be $D = 0.012\ um^2 \cdot s^{-1}$. This should be the case, since by using Einstein's relation (38) and considering the 20% increase in viscosity due to cholesterol [15] the theoretical value of the diffusivity is exactly $0.012\ um^2 \cdot s^{-1}$.

## V. Conclusion

A new model to explain Brownian yet non-Gaussian behavior is proposed, by including molecular interactions. Two regimes in a weakly interacting and highly compressible fluid are presented. Long-range interactions Brownian-particle fluid cause a phase transition for short distances. In the first regime the fluid behaves like a real gas interacting with an interface with no shear viscosity. If only Brownian particle-Brownian particle interaction via the fluid is considered, a double exponential distribution on the displacements of a Brownian particle is derived.

In the second regime, the fluid behaves as a liquid with non-zero viscosity, but with very high diffusivity (i.e., due to low compressibility/high mean free path caused by the separation of fluid particles via Lennard-Jones potential). Consequently, regular Brownian motion takes place and the double exponential transitions via a transition probability rapidly into a Gaussian. (All the pdfs are re-scaled, such that $\langle r(t)^2 \rangle$ is given by the 1D analogous of (32) since the Brownian particle might be in either the gaseous or in the liquid phase at some point in time).

The experimental data obtained by Wang *et al* is used to validate the model. It is shown that the theoretical model explains via a phase transition the physical cause for an unexpectedly high diffusivity. In addition, the lack of a phase transition when using cholesterol and the compressibility and shear deformation of the liquid (i.e., considering extensional viscosity) leads to a different and smaller value of the diffusivity which fully agrees with the experimentally reported value. Further, the pdfs for different times predicted by the model are compared with the experimental data and are found to be in excellent agreement. At $t_c = 4s$ the pdf starts transitioning away from exponential and after approximately $2s$ is found to be Gaussian, as reported in the experiment.

Finally, this approach links theoretical concepts in previous research of fluid behavior [26], [39] and [27] and theory of interfaces [30] to Brownian motion, which indicates that there are strong motives to expand studies of highly compressible liquids (in colloidal suspensions) and their molecular interactions with a Brownian particle (i.e.; different effective/interaction potentials to the one in (5) can be proposed to model different regimes at shorter time scales, where other diffusion processes may occur). Then, these effects can be incorporated into Langevin dynamics without the need of complicated mathematical modifications to noise terms.

## Data availability

Data and the code are available from the corresponding author on request.